\documentclass[aps,twocolumn,prl,superscriptaddress,showpacs]{revtex4-1}
\usepackage{amsmath}
\usepackage{amsfonts}
\usepackage{amssymb}
\usepackage{multirow}
\usepackage{verbatim}
\usepackage{alltt}
\usepackage{moreverb}
\usepackage{graphicx,color}
\usepackage{hyperref}
\usepackage{setspace}
\providecommand{\ignore}[1]{}
\usepackage{pdfpages}

\ignore{
tar czvf numsub_aps.tar.gz numsub_aps.tex aps_Figure{2,3b,3a,4}.jpg \
   aps_Figure4_inset_t.png
}

\newcommand{\aerrb}[2]{\protect\raisebox{1pt}{\protect\scalebox{.8}{\mbox{${}_{#1}^{#2}$}}}}
\begin{document}

\title{Generation of Optical Coherent State Superpositions by
  Number-Resolved Photon Subtraction from Squeezed Vacuum}

\author{Thomas Gerrits}
\affiliation{National Institute of Standards and Technology, Boulder, CO, 80305, USA}
\author{Scott Glancy}
\affiliation{National Institute of Standards and Technology, Boulder, CO, 80305, USA}
\author{Tracy S. Clement}
\affiliation{National Institute of Standards and Technology, Boulder, CO, 80305, USA}
\author{Brice Calkins}
\affiliation{National Institute of Standards and Technology, Boulder, CO, 80305, USA}
\author{Adriana E. Lita}
\affiliation{National Institute of Standards and Technology, Boulder, CO, 80305, USA}
\author{Aaron J. Miller}
\affiliation{Albion College, Albion, MI 49224, USA}
\author{Alan L. Migdall}
\affiliation{National Institute of Standards and Technology, Gaithersburg, MD, 20899, USA}
\affiliation{Joint Quantum Institute, Univ. of Maryland, College Park, MD 20742, USA}
\author{Sae Woo Nam}
\affiliation{National Institute of Standards and Technology, Boulder, CO, 80305, USA}
\author{Richard P. Mirin}
\affiliation{National Institute of Standards and Technology, Boulder, CO, 80305, USA}
\author{Emanuel Knill}
\affiliation{National Institute of Standards and Technology, Boulder, CO, 80305, USA}

\date{\today}

\begin{abstract}
We have created heralded coherent state superpositions (CSS), by
subtracting up to three photons from a pulse of squeezed vacuum
light. To produce such CSSs at a sufficient rate, we used our
high-efficiency photon-number-resolving transition edge sensor to
detect the subtracted photons. This is the first experiment enabled by
and utilizing the full photon-number-resolving capabilities of this
detector.  The CSS produced by three-photon subtraction had a mean
photon number of $2.75\aerrb{-0.24}{+0.06}$ and a fidelity of
$0.59\aerrb{-0.14}{+0.04}$ with an ideal CSS. This confirms that
subtracting more photons results in higher-amplitude CSSs.
\end{abstract}
\pacs{42.50.Dv, 42.50.Xa, 03.65.Ta, 03.65.Wj}

\maketitle

A coherent state of the electromagnetic field is often considered the
most classical-like pure state, but a superposition of two coherent
states with opposite phases has interesting quantum features.  For
example, coherent state superpositions (CSS) can be exploited for
performing quantum information tasks and high precision
measurements. CSSs are also of fundamental interest: When they contain
many photons they are superpositions of macroscopically
distinguishable states often called ``Schr\"odinger cat states''.
Schr\"odinger's Gedanken experiment of 1935 described a cat apparently
held in a superposition of alive and dead
states~\cite{schroedinger:qc1935a}, but many researchers now use
``Schr\"odinger cat'' to refer to a quantum state that is a
superposition of two highly distinguishable classical states such as a
CSS of high amplitude or mean number of
photons~\cite{yurke:qc1986a}. CSSs have been prepared in traveling
optical modes with a mean of up to 2.0 optical photons by
heralding~\cite{ourjoumtsev:qc2006b,neergaard-nielsen:qc2006a,wakui:qc2007a,ourjoumtsev:qc2007b,takahashi:qc2008a}.
With sufficiently high quality and well characterized CSSs, one can in
principle quantum compute using simple linear optical components and
homodyne measurements~\cite{gilchrist:qc2003a}. Less ambitiously, they
can serve as flying qubits for quantum communication.  In addition to
potentially simple processing, advantages of CSSs in traveling optical
modes include fast linear manipulations, transport over large
distances, robustness if loss is controlled, and simple conversion to
entangled optical states, all at room temperature.

The CSSs that we discuss here are superpositions of two coherent
states $|\pm \alpha\rangle$ of a single mode of light, where $+\alpha$
and $-\alpha$ are the states' complex mode amplitudes. Our experiments
aim to prepare two special instances of these CSSs: the odd and even
CSSs defined as the superpositions $|{-}\alpha\rangle \pm
|\alpha\rangle$ (unnormalized). These are distinguished by having only
even ($+$) or odd ($-$) numbers of photons. For $|\alpha|\gg 1$, the
states' mean number of photons, $\langle n\rangle$, is approximately
$|\alpha|^2$. Two quality measures for experimental CSSs are the
fidelity of the created state with the nearest ideal CSS and the
magnitude of the amplitude of this ideal CSS. There are two reasons to
aim for large amplitude CSSs. The first is that to be useful for
superresolution metrology, the probability $p_0=1-\exp(-2|\alpha|^2)$
with which the superposed coherent states can be distinguished must be
close to one. To achieve $p_0>0.99$ requires $|\alpha|>1.52$. The
second is that a minimum size estimated as $|\alpha|>1.2$ is required
for fault tolerant quantum computing~\cite{lund:qc2008a}. Because
operation close to the lower bound is unrealistic due to excessive
resource requirements, we are motivated to produce bigger
CSSs. Similarly, high fidelity is required to avoid excessive
overheads for eliminating unwanted errors due to deviations from an
ideal CSS. The highest fidelity CSS achieved so far has $|\alpha|=1.1$
and a fidelity $F=0.76$~\cite{takahashi:qc2008a}, while the largest
has an effective size of $|\alpha|=1.4$ and fidelity
$F=0.60$~\cite{takahashi:qc2008a}. We have created CSSs with
amplitudes and fidelities of 
$|\alpha|=1.76\aerrb{-0.19}{+0.02}$ 
and
$F=0.59\aerrb{-0.14}{+0.04}$, and $|\alpha|=1.32\aerrb{-0.02}{+0.01}$ and
$F=0.522\aerrb{-0.010}{+0.004}$. Unlike the experiment reported in
\cite{takahashi:qc2008a}, our CSSs are generated in pulsed rather than
continuous-wave mode. Pulsed operation is required for many
applications to avoid the effects of light in neighboring modes in
subsequent manipulations and measurements of the states.

\begin{figure}
\includegraphics[width=3.25in,height=1.84in]{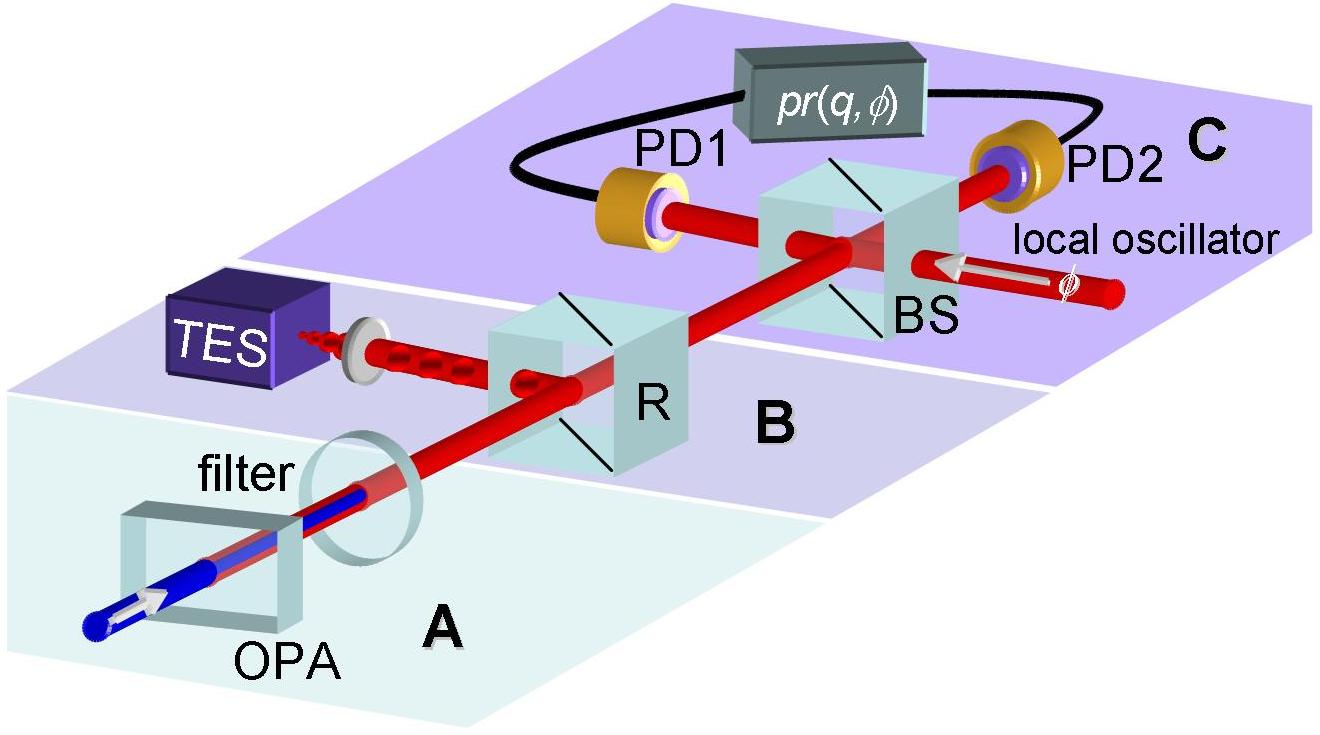}
\caption{(color online) Scheme for optical coherent state
  superposition (CSS) creation. An upconverted laser pulse enters an optical
  parametric amplifier (OPA) to create a squeezed vacuum state in
  section (A). After spectral filtering, this state is sent to a
  weakly reflecting beamsplitter R in (B). Reflected photons that are
  detected herald a CSS emerging from R into (C).  Its quadratures are
  measured by homodyne detection in (C).}
\label{fig1}
\end{figure}

To create the CSSs, we used the photon subtraction scheme depicted in
Fig.~\ref{fig1}. A squeezed vacuum state is prepared and sent through
a weakly reflecting beam splitter. Reflected photons that are detected
herald an approximate CSS in the transmitted beam.  Because higher
amplitude and fidelity CSSs can be created by heralding on detecting
multiple photons at once~\cite{dakna:qc1997a,glancy:qc2008a}, we used
a photon-number-resolving transition edge sensor
(TES)~\cite{lita:qc2008b,rosenberg:qc2007a} for subtracting two or
three photons. The TES used in our experiment has an efficiency of
$85\pm 2\;\%$ and can resolve up to 10 photons.  This enabled
obtaining higher amplitude CSS at practical rates.  We also subtracted
one and two photons using avalanche photodiodes (APDs) for comparison.

For the experiments, we used a cavity-dumped $861.8\;\textrm{nm}$
laser with transform-limited pulses of $140\;\textrm{fs}$ (typical),
pulse energies of $40\;\textrm{nJ}$ and a repetition frequency of
$548\;\textrm{kHz}$. A fraction of each pulse with $>10^9$ photons was
used as the local oscillator (LO) in the homodyne detector. The rest
pumped a temperature-tuned $150\;\textrm{$\mu$m}$ KNbO${}_3$ crystal
to generate a second-harmonic pump pulse (efficiency $25\;\%$) for the
optical parametric amplifier (OPA) shown in Fig.~\ref{fig1}.  
The OPA consists of a temperature-tuned
$200\;\textrm{$\mu$m}$ long  KNbO${}_3$ crystal. We determined that the
squeezed vacuum state generated can be modeled as a pure squeezed
state with minimum quadrature variance $V_0=-6.8\;\textrm{dB}$
subjected to a loss of $\gamma_s=0.36$. We define the squeezing purity
as $\eta_s=1-\gamma_s$.  We used a variable beam splitter (R in
Fig.~\ref{fig1}) made with a half-wave-plate and a polarizing
beamsplitter and configured to send from $2.5\;\%$ (one-photon
subtraction) to $20\;\%$ (three-photon subtraction) of the light to
the photon subtraction arm. Photons in this arm were spectrally
filtered by a fiber Bragg grating with a bandwidth of
$1.5\;\textrm{nm}$ in a polarization-based circulator before being
coupled to the photon detector/counter. The other arm of the variable
beam splitter delivers the heralded CSS to a conventional homodyne
detector for measuring the quadrature at the phase of the
LO. The CSS temporal shape is significantly different from that of the
original pump due to the large mismatch in group velocity in our
KNbO${}_3$ crystals. To compensate, we expanded the temporal width of
the LO by up to a factor of $2$ with a pulse-shaping
setup~\cite{weiner:qc2000a}. The phase of the LO was adjusted by a
piezo-mounted mirror displaced at a frequency of
$2.75\;\textrm{Hz}$ with a saw-tooth profile to obtain a complete
phase space measurement of the CSS.  Further technical details
are in~\cite{epaps}.

We reconstructed the states produced by photon subtraction immediately
after the subtracting beam splitter by maximum likelihood quantum
state estimation as discussed in Ref.~\cite{lvovsky:qc2004a}. For this
purpose, we considered the homodyne measurement setup including all of
its losses such as those associated with the initial beamsplitter and
imperfect spatial mode matching to the LO as a monolithic lossy
quadrature measurement. This requires knowing the loss $\gamma_h$,
which we experimentally determined to be $\gamma_h=15\pm2\;\%$. The
uncertainty in $\gamma_h$ propagates to an uncertainty in the reported
CSS parameters.  In particular, the fidelities differ by up to $\pm
0.02$ if the boundary values for $\gamma_h$ are used. However, the
main uncertainty in our state reconstructions is due to finite sample
statistics. We estimated this statistical uncertainty by
parametric-bootstrap resampling~\cite{efron:qc1993a}.  We report
inferred values such as fidelities in the form $F_{{-(F-L)}}^{{U-F}}$,
where $F$ is the fidelity of the maximum likelihood estimate from the
experiment's data, $U$ is the $84^\textrm{th}$ percentile of the
fidelities of the states estimated from resampled data sets, and $L$
is the $16^\textrm{th}$ percentile. We obtained $100$ resampled data
sets for one- and two-photon subtraction and $1000$ for three-photon
subtraction. There is a significant bias toward more mixed states in
the resampling procedure and the amount of bias increases with the
purity of the state from which samples are generated. We did not
correct for this bias in our reconstruction of the states, but note
that it suggests that the true fidelities are above the reported ones.

The reconstructed states have well-defined average photon numbers,
$\langle n \rangle$. The reported amplitudes are those of the nearest
even or odd CSS, which is found by maximizing the fidelity with
respect to the reconstructed state. The reported fidelities are these
maximized ones.  Table~\ref{table1} summarizes our results.

\begin{figure}
\includegraphics[height=2.6in]{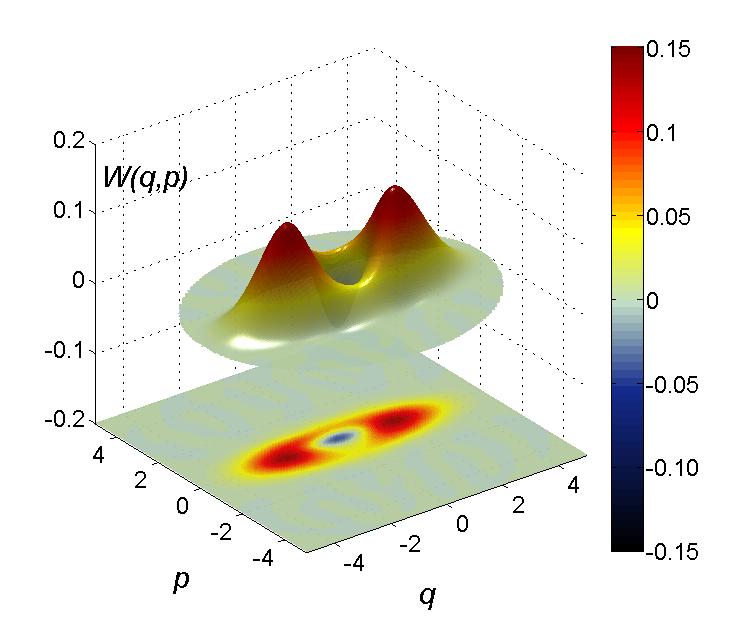} 
\caption{(color online) Maximum likelihood estimate of an odd CSS
  generated by one-photon subtraction from a squeezed vacuum. The
  graph shows the unitless Wigner function value $W(q,p)$ as a
  function of the unitless quadratures of the electromagnetic field. }
\label{fig2}
\end{figure}

 \begin{table}[h]
   \begin{center}
    \caption{Results for the one-, two- and three-photon subtraction experiments. $W_{\textrm{min}}$ and$\langle n \rangle$ are the minimum value and the mean photon number of the reconstructed state, respectively. $F$ is the fidelity of the reconstructed state compared to a theoretical CSS with amplitude $|\alpha|$}
   \begin{tabular}{l c c c c}
    \hline
    \hline
    $ $ & $W_\textrm{min}$ & $\langle n \rangle$ & $F$ & $|\alpha|$\\
    \hline
    \multicolumn{5}{l}{One-photon experiment:}\\
    $ \textrm{APD}$ & $-0.041\aerrb{-0.001}{+0.009}$ & $1.96\aerrb{-0.04}{+0.05}$ & 
    $0.522\aerrb{-0.010}{+0.004}$ & $1.32\aerrb{-0.02}{+0.01}$\\[3pt]
    {Ref.}~\cite{ourjoumtsev:qc2006b} & $-0.13\pm 0.01$ & $$ & $0.70$ & $0.89$\\
    \hline
    \multicolumn{5}{l}{Two-photon experiments:}\\
    $\textrm{APDs}$ & $-0.018\aerrb{-0.002}{+0.002}$ & $2.34\aerrb{-0.05}{+0.06}$ & 
    $0.523\aerrb{-0.014}{+0.022}$ & $1.30\aerrb{-0.02}{+0.04}$
\\[3pt]
    $\textrm{TES}$ & $-0.010\aerrb{-0.001}{+0.001}$ & $1.89\aerrb{-0.06}{+0.05}$ & 
    $0.531\aerrb{-0.018}{+0.017}$ & $1.16\aerrb{-0.04}{+0.04}$\\

    {Ref.}~\cite{takahashi:qc2008a} & $$ & $$ & $0.60$ & $1.4$\\
    \hline
    \multicolumn{5}{l}{Three-photon experiment:}\\
    $\textrm{TES}$ & $-0.116\aerrb{-0.019}{+0.073}$ & $2.75\aerrb{-0.24}{+0.06}$ & 
    $0.59\aerrb{-0.14}{+0.04}$ & $1.76\aerrb{-0.19}{+0.02}$\\
    \hline
   \end{tabular}
    \label{table1}
   \end{center}
   \end{table}

Fig.~\ref{fig2} shows the reconstructed Wigner function for a
one-photon-subtracted state heralded by an APD.  The quantum character of
this state can be identified by its negativity near the origin of the
Wigner function, whose minimum has a value of
$W_\textrm{min}=-0.041\aerrb{-0.001}{+0.009}$.  The state's fidelity
is $F=0.522\aerrb{-0.010}{+0.004}$ with respect to an odd CSS with
$|\alpha|= 1.32\aerrb{-0.02}{+0.01}$.  This fidelity is higher than
the maximum fidelity of $F=0.487$ that any coherent state can have
with the $|\alpha|=1.32$ odd CSS. (Note that this is also the highest fidelity
that any mixture of coherent states can have. The maximum fidelity of
a coherent state with a CSS depends on the CSS's $|\alpha|$ and
whether the CSS is even or odd.  As $|\alpha|$ increases, this
fidelity approaches 0.5 from above for even CSSs but from below for
odd CSSs.) The amplitude of the CSS is notably larger than the
$|\alpha|=0.88, F=0.70, W_\textrm{min}=-0.13$ state described in
Ref.~\cite{ourjoumtsev:qc2006b}.  The lower fidelity is primarily due
to a lower squeezing purity $\eta_s$ in our
experiment.

\begin{figure}
\begin{picture}(3.2,1.7)(0,-1.7)
\put(3.35,0){\makebox(0,0)[tr]{
    \includegraphics[height=1.62in]{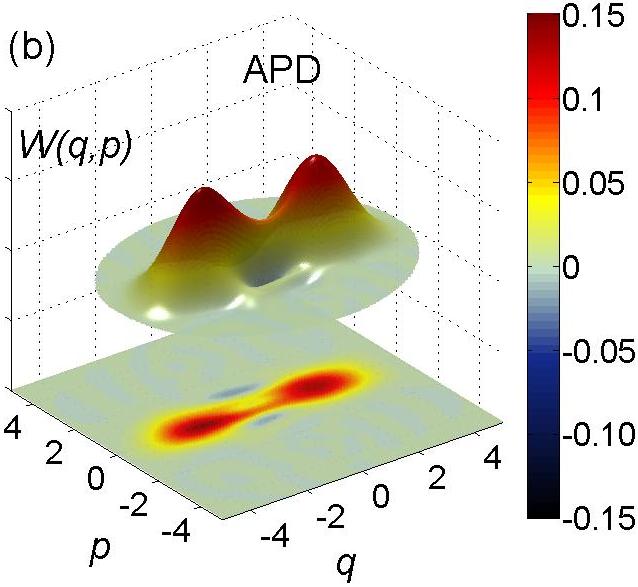} 
}}
\put(-0.15,0){\makebox(0,0)[tl]{
    \includegraphics[height=1.62in]{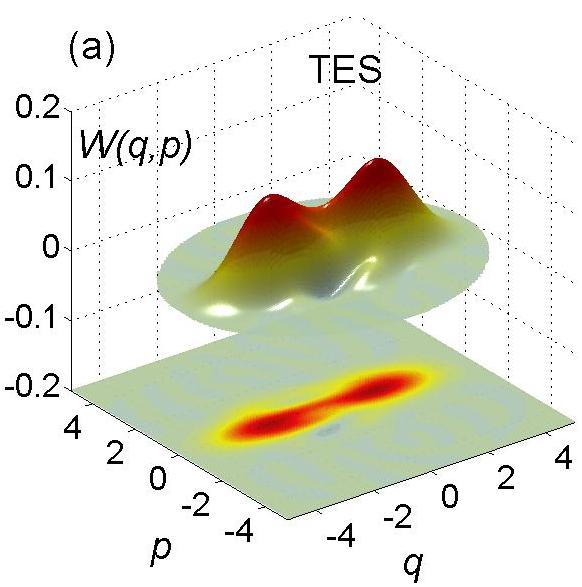} 
}}
\end{picture}
\caption{(color online) Wigner functions of the maximum likelihood
  estimates of even CSS created by two-photon subtraction and heralded
  with (a) one transition edge sensor and (b) two multiplexed APDs.}
\label{fig3}
\end{figure}

We obtained even CSSs by two photon subtraction. We performed two
experiments, the first used a TES, the second used two APDs at the two
outputs of a $50/50$ beamsplitter. For the APDs, coincidence heralds the
presence of two photons in the subtraction arm. The reconstructed
states are shown in Fig.~\ref{fig3}.  The TES measurement yielded a
smaller CSS ($|\alpha_{\textrm{TES}}|=1.16\aerrb{-0.04}{+0.04}$) than the
APD measurement ($|\alpha_{\textrm{APD}}|=1.30\aerrb{-0.02}{+0.04}$). The
fidelities are $F_\textrm{TES}=0.531\aerrb{-0.018}{+0.017}$ and
$F_\textrm{APD}=0.523\aerrb{-0.014}{+0.022}$.  For comparison, the maximum
fidelity of coherent states with an $|\alpha|=1.16$ ($|\alpha|=1.30$)
even CSS is $0.552$ ($0.522$). Earlier
studies~\cite{takahashi:qc2008a} showed the continuous wave generation
of even CSSs with $|\alpha|=1.41$ and $F = 0.60$.

The fidelity of the heralded CSSs is affected not only by low
squeezing purity, but also by unwanted photons not matching the LO
mode but still visible to the detectors. In addition to stray light
(which can in principle be controlled) such photons come from
temporally similar modes that are also squeezed in the OPA.  When
squeezed light is produced by down-conversion of a pulsed pump laser,
multiple spatial-temporal modes may be squeezed, and none of these
modes is guaranteed to match the mode of the
LO~\cite{wasilewski:qc2005a}. These other modes have similar spectra
to the LO mode and therefore cannot be conventionally
filtered. Detections due to photons in these modes degrade the
fidelity of the CSSs.  We quantify the effect of unwanted photons with
the ``modal purity'' $\xi_n$ of $n$ photon subtraction -- the
probability that, when the subtraction detector registers $n$ photons,
these $n$ photons were from the mode matching the LO. To estimate the
modal purities, we used a single-mode photon subtraction model to fit
our data~\cite{epaps}. From this we determined
$\xi_{2,\textrm{TES}}=0.62$ and $\xi_{2,\textrm{APD}}=0.85$, compared
to $\xi_{1}=0.91$ for the one-photon subtraction experiment. The
reason for the lower modal purity of the TES experiment is its greater
sensitivity to stray photons from the LO. With the APDs, we can gate
the detections to reject slightly delayed LO photons arising from
downstream reflections. The TES is slower, so such gating is not
possible.

The main advantages of the TES are the greater efficiency and the
ability to directly count photons. In the two-photon subtraction
experiments, this higher efficiency resulted in improving the rate at
which CSSs were heralded by a factor of three.

\begin{figure}
    \includegraphics[height=3.1in]{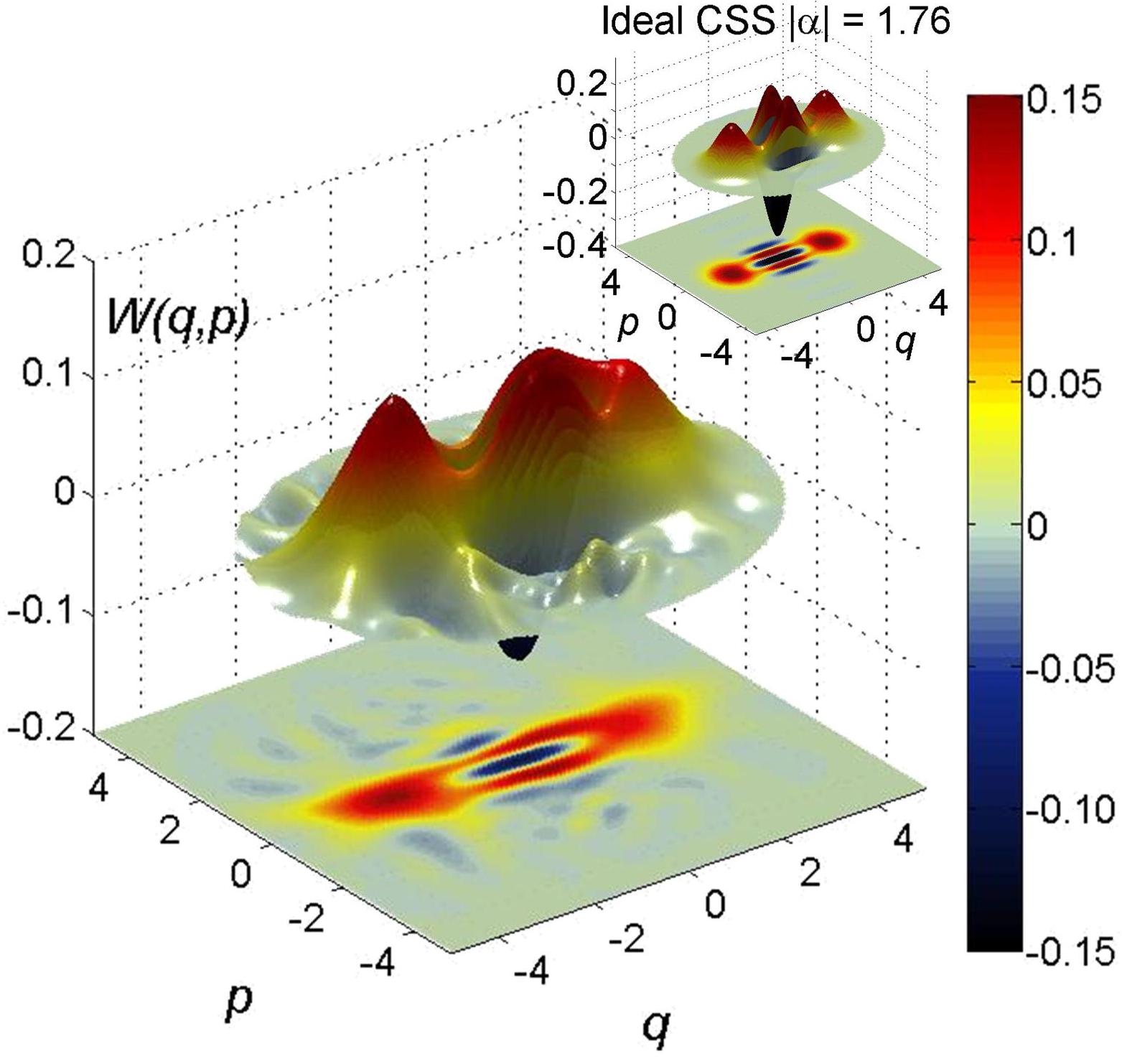} 
\caption{(color online) Maximum likelihood estimate of an odd CSS
  after the subtraction of three photons from a squeezed vacuum. The
  reconstructed state has a fidelity of $F =
  0.592\aerrb{-0.142}{+0.036}$ with a CSS of amplitude $|\alpha| =
  1.76\aerrb{-0.19}{+0.02}$. Inset: Wigner function of an ideal odd
  CSS with $|\alpha|=1.76$}
\label{fig4}
\end{figure}

Three-photon subtraction events are extremely rare in our experiment.
Nevertheless, using the TES we were able to detect $1087$ three photon
events over a period of approximately $60$ hours. With three
multiplexed APDs we would have collected only about $120$
events. Fig.~\ref{fig4} shows the odd CSS. To increase the
three photon event rate, we increased the reflectivity of the photon
subtraction beam splitter to $20\;\%$, sacrificing the fidelity of the
CSS. The reconstructed state shows a negative minimum of its Wigner
function $W_\textrm{min} = -0.116\aerrb{-0.019}{+0.073}$ and a mean photon
number of $2.75\aerrb{-0.24}{+0.06}$.  The state has fidelity
$F=0.59\aerrb{-0.14}{+0.04}$
with an ideal CSS of $|\alpha|=1.76\aerrb{-0.19}{+0.02}$. The
estimated modal purity in this experiment is $\xi_3=0.84$.  Thus, we
observed the predicted increase in CSS amplitude for three-photon
subtraction, but the increase in fidelity is not statistically
significant.

In conclusion, we have measured heralded optical CSSs created by
subtracting up to three photons from a squeezed vacuum state, using
APDs for one- and two-photon subtraction and a TES for two and three.
It was only by taking advantage of the high efficiency and
the direct photon counting capability of the TES that we were able to
successfully subtract three photons with a sufficiently high rate of
CSS production. The CSSs produced were analyzed by homodyne
measurement and maximum-likelihood state estimation.  The quality
of the CSSs can be improved by reducing the losses experienced by the
squeezed vacuum state before reaching the photon-subtraction beam
splitter. For multi-photon subtraction, however it is crucial to reduce
the presence of unfilterable photons in unwanted modes.  A promising
route that addresses both problems is to tailor the squeezing source
to create squeezed light only in a single mode matched to the LO.
This route is being pursued in the photon-pair generation
community~\cite{grice:qc2001a,mosley:qc2008a,avenhaus:qc2009a}.  Based
on our findings, we propose that the combination of pure vacuum
squeezing and high efficiency detectors with photon-number-resolving
capabilities can yield high rate, amplitude and fidelity CSSs to
support quantum information processing and metrology beyond the
quantum limit.
\\

Added note: Recently, the authors became aware of a similar measurement 
that made use of  photon-number-resolving transition edge sensors~\cite{Namekata:2010}.
\\

\begin{acknowledgments}
This work was supported by the NIST Innovations in Measurement Science
Program. T.G. thanks P. Grangier and A. Ourjoumtsev for
discussions. This is a contribution of NIST, an agency of the
U.S. government, not subject to copyright.
\end{acknowledgments}

\clearpage


\section*{Supplementary material}

# Generation of Optical Coherent State Superpositions by Number-Resolved Photon Subtraction from Squeezed Vacuum *

## Supplementary Material

Thomas Gerrits[1], Scott Glancy[1], Tracy S. Clement[1], Brice Calkins[1], Adriana E. Lita[1], Aaron J. Miller[3], Alan L. Migdall[2], Sae Woo Nam[1], Richard P. Mirin[1], Emanuel Knill[1]

[1]*National Institute of Standards and Technology, Boulder, CO, 80305, USA*
[2]*National Institute of Standards and Technology, Gaithersburg, MD, 20899, USA and Joint Quantum Institute, Univ. of Maryland, College Park, MD 20742, USA*
[3]*Albion College, Albion, MI 49224, USA*

This supplementary material gives more technical experimental details and deeper insight into the analysis than was given in the main paper. First, we give a detailed description of the experimental setup and our efforts to improve the fidelity of our squeezed vacuum and coherent state superpositions (CSSs). Then we present an analytical model of subtraction of up to three photons from a squeezed vacuum. Last we give a detailed analysis of experimental parameters and results.

## I. Experimental Setup

Figure A1 shows our experimental setup. We use a cavity-dumped femtosecond laser with transform-limited pulses of typically 140 fs duration and a repetition frequency of 548 kHz. The center wavelength of the cavity dumper output is $\lambda_0 = 861.8$ nm. Typical pulse energies are 40 nJ at the output port of the cavity dumper. We spatially filter the laser beam by sending it through a 30 µm diameter pinhole (PH). At the 90/10 beam splitter (BS) we split the laser beam into two parts. The weaker part of the beam is the strong local oscillator (LO) for the homodyne detection ($>10^9$ photons/pulse). The stronger part pumps a 150 µm thick $KNbO_3$ crystal (SHG) to generate the second-harmonic pump photons with a conversion efficiency of 25%. To eliminate any fundamental photons from the laser itself, we spectrally filter (SF) the second-harmonic pump. Then, the pump is focused into a 200 µm thick down-converting $KNbO_3$ crystal (OPA). Both crystals are temperature-tuned with stability better than 0.05°C for optimum phasematching. We optimized both crystal temperatures to achieve purest squeezing ($T_{OPA} = 28.3°C$; $T_{SHG} = 27.5°C$). During the course of one measurement we observe a decrease in the squeezing from the OPA, which is probably caused by a photorefractive effect present in our crystals [1]. This decrease typically happens within the first hour of the measurement and remains stable afterwards if we keep the pump focused at the same position. Our optics in the homodyne detection arm after the down-converter eliminate most ($>99.9\%$) of the pump, and its contribution to the homodyne signal is negligible. Therefore, no further spectral filtering is required after the OPA.

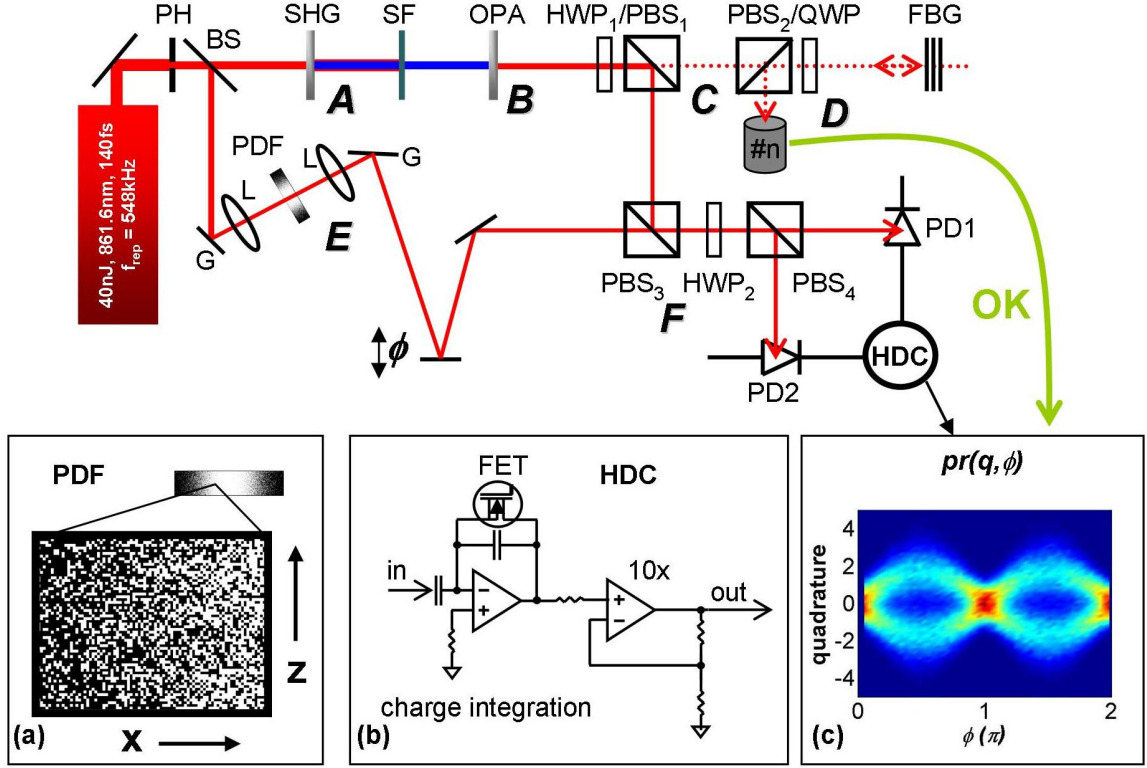

FIG. A1. Experimental Setup. The experiment setup consists of six main sections, labeled A-F. **A**: second-harmonic conversion of fundamental beam; **B**: squeezed vacuum generation; **C**: photon subtraction; **D**: spectral filtering of subtracted photons and detection thereof; **E**: temporal pulse-shaping of local oscillator; **F**: homodyne detection. The inset (a) shows the polka dot filter used in the pulse-shaping setup. Inset (b) shows the charge-integration circuit for homodyne detection, and inset (c) shows quadrature data for a full scan of the phase of a CSS created by one photon subtraction. PH: pinhole; BS: beam splitter; SHG: second-harmonic generation; SF: spectral filter; OPA: down-conversion crystal; HWP: half-wave-plate; PBS: polarizing beam splitter; QWP: quarter-wave-plate; FBG: fiber Bragg grating; #n: photon-number-resolving detector; PD: photodiode; HDC: homodyne detection circuit; $\phi$: Piezo stage; G: grating; L: lens; PDF: polka dot filter

The squeezed vacuum is sent to our photon-subtraction beam-splitting components. The beam splitter consists of a half-wave-plate and a polarizing beam splitter cube (HWP$_1$/PBS$_1$). This arrangement is helpful, as we can adjust the beam splitter's transmissivity, rather than having a fixed splitting ratio. The s-polarized waves experience less than 1% loss under reflection and the p-polarized waves undergo a 5% loss due to transmission through the polarizing beam splitter. The reflected photons are sent to the homodyne detection arm, and hence the loss in this arm is minimized. The subtracted photons are directed to our spectral filter setup. It consists of a fiber Bragg grating (FBG) and a circulator. The FBG has a bandwidth of $\Delta\lambda_{\mathrm{FWMH}} = 1.5$ nm. The circulator is a combination of a free-space polarizing beam splitter and quarter-wave-plate (PBS$_2$/QWP). This setup allows for good filtering of the subtracted photons. However, two fiber couplers are in the path of the subtracted photons. This decreases the overall detection efficiency of the subtracted photons due to the limited fiber coupling efficiency. The subtracted photons that are within the FBG bandwidth reach the photon-

number-resolving detector with a probability of about 20%. We used two different photon detectors: avalanche photo diodes (APDs) for the one and two photon subtraction, or one transition-edge sensor (TES) for the two and three photon subtraction experiments. Upon detection of at least one photon at the photon detector, we know that we have prepared an approximation of a CSS. Homodyne measurements are recorded regardless of whether a photon is subtracted or not. When no photons are subtracted, a noisy squeezed state is created, which we use to calibrate the phase of the LO. When a photon subtraction event occurs, the coincident homodyne measurement is tagged.

The temporal width of the strong LO is controlled by a pulse-shaping setup [2] so that we can compensate for the large mismatch in group velocity in our KNbO$_3$ crystals, which is ~1.2 ps/mm [3]. This group velocity mismatch causes the temporal width of the squeezing pulse to be about double that of the LO. Using two gratings (G), two lenses (L) and a spatial filter (PDF), we can tune the temporal width of the LO from 140 fs to about 300 fs. The spatial filter is a polka dot filter (PDF) with Gaussian transmission profile to minimize possible chirp imposed by the pulse-shaping setup. We lithographically made these polka dot filters on chromium masks. In order to minimize and randomize the interference of the LO, which was sent through the PDF, we randomly distributed 4 μm × 4 μm squares along the z-axis of the filter. The squares' density changed according to a Gaussian envelope transmission profile along the x-axis of the filter. The filter design is shown in inset (a) of figure A1. Different LO temporal widths are achieved by different PDF designs. The highest fidelity CSS results were obtained for an LO width of 230 fs. However, this is about 40 fs shorter than the width that results in the most pure squeezed states, where the squeezing is measured by direct homodyne detection. The phase of the LO is adjusted by a piezo-mounted mirror ($\phi$). We continuously displace the mirror with a frequency of 2.75 Hz. The mirror's displacement is a saw-tooth profile with amplitude of about $2\lambda_0$. This allows a complete phase space measurement of our CSS. The LO and prepared CSS are combined at PBS$_3$. A half-wave-plate (HWP$_2$) and a polarizing beam splitter (PBS$_4$) constitute the 50/50 beam splitter necessary for the homodyne detection setup. Adjustment of HWP$_2$ allows for very accurate balancing of the homodyne detection system. Both photodiodes are high-speed pin-Si photodiodes with high detection efficiency. Out of a set of 10 same-wafer photodiodes, we chose the two photodiodes that have the best matching temporal response. The electrical homodyne circuit is shown in inset (b) of figure A1. It consists of a charge integration stage and a 10× amplification stage. The charge integration includes one field effect transistor (FET) bridged across the integrating capacitor. The FET circuit induces a random offset charge. We correct for this charge by subtracting the voltage before from the voltage after each laser pulse. An intensity plot of a quadrature probability distribution for a one photon subtraction experiment is shown in inset (c). The quadrature data consist of 324,000 heralded events. These data are then processed, and maximum likelihood estimation gives the density matrix of the measured state.

The overall efficiency of the homodyne detection is $\eta_h = 0.853 \pm 0.028$. We estimate this by separately measuring four efficiencies $\eta_o$, $\eta_d$, $\eta_w$, and $\eta_e$ and computing $\eta_h = \eta_o \eta_d \eta_w \eta_e$. The four efficiencies are defined as follows: 1) The efficiency of all optical elements after the photon subtraction beam splitter to the face of the homodyne detector's photodiodes is $\eta_o = 0.94 \pm 0.005$. 2) The mean efficiency of the photodiodes is $\eta_d = 0.976 \pm 0.022$, where the difference in efficiency between the two photodiodes is

less than 0.5%. 3) The efficiency of the mode matching between the squeezed mode and the LO is $\eta_w = 0.95 \pm 0.005$. $\eta_w$ was determined by interfering a probe beam with the LO. However, the probe beam does not travel through the up-conversion crystal and its temporal width has not been altered. Therefore, for this measurement we did not alter the LO's temporal width. The LO is simply sent through the pulse shaper without a polka dot filter in place. (We are unable to measure the mismatch between the temporal width of the CSS and the LO, so it is not included in $\eta_h$.) The error bar for $\eta_w$ accounts only for the statistics of the measurement, not for likely systematic biases. The effect of such biases on the inferred states is well below the states' statistical error as discussed in the paper. 4) $\eta_e$ is the efficiency that is formally equivalent to the electronic background noise. The electrical background noise of the homodyne detectors and electronics is $e = V_e/(V_e + V_v) = 0.021 \pm 0.001$, where $V_e$ is the variance of voltages measured when no light enters the photodiodes and $V_e + V_v$ is the variance observed when only the local oscillator is present ($V_e + V_v$ includes both electronic noise and shot noise of the LO). This is formally equivalent to an efficiency $\eta_e = (1\text{-}e) = 0.979 \pm 0.001$ [4]. Our calibration of the overall efficiency of the homodyne detection, $\eta_h$, does not include any excess noise from the down-conversion process itself, nor does it include the reflectivity of the beam splitter used for photon subtraction.

We model the squeezed vacuum generated in the experiment as a pure squeezed vacuum state with squeezed quadrature variance $V_0$, which has passed through a medium with transmissivity $\eta_s$. This state is then measured with the homodyne system whose efficiency is $\eta_h$; hence the measured efficiency is $\eta_m = \eta_h\eta_s$. (When measuring the squeezing directly, we set the photon subtraction beam splitter reflectivity to 0.) We can calculate $\eta_m$ and $V_0$ by use of [5]:

$$\left(1-V_p\right) = \eta_m\left(1-V_0\right),\tag{A1}$$

and

$$\eta_m = \frac{(1-V_q)(1-V_p)}{2-V_q-V_p},\tag{A2}$$

where $V_p$ is the observed squeezed quadrature's variance, and $V_q$ is the observed anti-squeezed quadrature's variance. When we measured the squeezing by homodyne detection directly, we observed $V_q = 3.129$ (+5.0 dB) and $V_p = 0.565$ (-2.5 dB). After correcting for the overall homodyne detection efficiency $\eta_h$, we found $\eta_s = 0.64$ and an inferred squeezing variance of $V_0 = 0.205$ (-6.8 dB), based on equations A1 and A2.

## II. Photon subtraction from squeezed vacuum (model calculations)

In the following, we present an analytical model of the photon subtracted squeezed states. The model is similar to the model published by Ourjoumtsev *et al.*[1]. However, it considers up to three photons subtracted from the squeezed vacuum and does not assume that the efficiency of the photon subtraction detector is small.

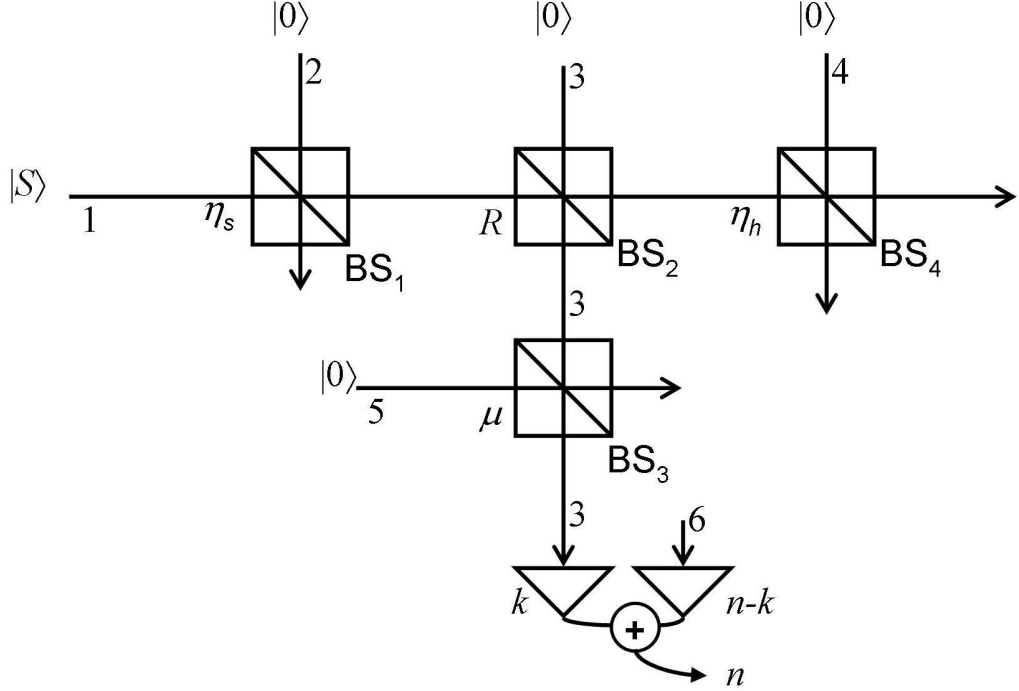

FIG. A2. Schematic of the model for calculation of the two-mode Wigner function. A squeezed state is sent into a beam splitter (BS$_1$), with transmissivity $\eta_s$, to model the loss of the input noisy squeezed state. BS$_2$ models the photon subtraction at a beam splitter with reflectivity $R$. BS$_3$ represents the loss in the photon subtraction mode given by the subtraction arm efficiency ($\mu$). The two triangles represent ideal photon counters, one of which detects $k$ photons in mode 3, and the other detects $n - k$ photons in orthogonal modes (all of which we label as mode 6). BS$_4$ models the known optical efficiency in the homodyne subtraction arm $\eta_h$, which includes the optical loss, photodiode efficiency, wavefront overlap and electrical background.

Unlike the model in [6], we do not consider multimode interference effects – we treat photons in modes not matched to the local oscillator as if they are equivalent to dark counts in the subtraction detector. The APDs are unable to distinguish the $n$ photon subtraction events from $n + 1$ or higher numbers, so in this experiment and our model we accept a heralding of an $n$ photon subtraction event when the detector registers $n$ or more photons. Although the TES detectors can distinguish $n$ from $n + 1$ photons we use the same model to analyze the TES experiments. Because the frequency of the $n + 1$ events is so much smaller than the $n$ photon events, their contribution is negligible. Figure A2 shows a schematic for the model. Beginning with the pure squeezed state $|S\rangle$ (with squeezed variance $V_0$) in mode 1 and vacuum in modes 2, 3, 4, and 5 we calculate the state of this system as it evolves through the beam splitters, trace out the lost modes, and project mode 3 onto a $k$ photon Fock state. We will account for the effect of dark counts in the subtraction detectors and photons in other modes entering the subtraction detector by including a second virtual photon counter, which registers $n - k$ photons, so that the total number of photons detected is $n$.

Initially, this system is in the state

$$\rho_{1,2,3,4,5} = |S\rangle_1 |0\rangle_2 |0\rangle_3 |0\rangle_4 |0\rangle_5 {}_1\langle S|{}_2\langle 0|{}_3\langle 0|{}_4\langle 0|{}_5\langle 0|. \qquad (A4)$$

We apply beam splitters 1 through 4 to this system and then perform the partial trace over modes 2, 4, and 5, leaving the state of modes 1 and 3:

$$\rho_{1,3} = \mathrm{Tr}_{2,4,5}\left[\hat{B}_{1,4}(\eta_h)\hat{B}_{3,5}(\mu)\hat{B}_{1,3}(1-R)\hat{B}_{1,2}(\eta_s)\rho_{1,2,3,4,5}\hat{B}_{1,2}^{\dagger}(\eta_s)\hat{B}_{1,3}^{\dagger}(1-R)\hat{B}_{3,5}^{\dagger}(\mu)\hat{B}_{1,4}^{\dagger}(\eta_h)\right]. \qquad (A5)$$

Next we project mode 3 onto the $k$ photon Fock state

$$\rho_1(k) = \frac{{}_3\langle k|\rho_{1,3}|k\rangle_3}{P(k)}, \qquad (A6)$$

where $P(k)=\mathrm{Tr}[{}_3\langle k|\rho_{1,3}|k\rangle_3]$ is the probability that $\rho_{1,3}$ contains $k$ photons in mode 3. Suppose mode 6 contains $x$ photons with probability $q(x)$. The probability that the two detectors together register $n$ photons is

$$Q(n) = \sum_{k=0}^{n} q(n-k)P(k), \qquad (A7)$$

and the resulting state is

$$\sigma_1(n) = \frac{1}{Q(n)}\sum_{k=0}^{n} q(n-k)P(k)\rho_1(k). \qquad (A8)$$

We accept a heralding of an $m$ photon subtracted state whenever $m$ or more photons are detected, so we consider the statistical mixture of all $\sigma_1(n)$ for $n \geq m$:

$$\tau_1(m) = \frac{1}{S(m)}\sum_{n=m}^{\infty} Q(n)\sigma_1(n), \qquad (A9)$$

where the probability to detect at least $m$ photons is

$$S(m) = \sum_{n=m}^{\infty} Q(n) = 1 - \sum_{n=0}^{m-1} Q(n). \qquad (A10)$$

In fact $\tau_1(m)$ does not depend on all of the details of the distribution $q(x)$, and we simplify these expressions using modal coupling parameters $\xi_n$, for all $n$ such that $n$ is an integer, and $0 \leq n \leq m$. We define the $\xi_n$ as

$$\xi_n = \begin{cases} \dfrac{1}{S(m)}\left(1 - \displaystyle\sum_{s=0}^{m-1} q(s)\right) & \text{for} \quad n = 0, \\[2ex] \dfrac{q(m-n)}{S(m)}\left(1 - \displaystyle\sum_{s=0}^{n-1} P(s)\right) & \text{for} \quad 1 \le n \le m. \end{cases} \qquad (A11)$$

Note that $\sum_{n=0}^{m} \xi_n = 1$, and the modal coupling parameters can be interpreted as probabilities. Using the modal coupling parameters we can rewrite $\tau_1(m)$ as

$$\tau_1(m) = \sum_{n=0}^{m} \xi_n\left(\sum_{k=n}^{\infty} P(k)\rho_1(k)\right) = \sum_{n=0}^{m} \xi_n\left(\rho_{id} - \sum_{k=0}^{n-1} P(k)\rho_1(k)\right), \qquad (A12)$$

where $\rho_{id}=\text{Tr}_3(\rho_{13})$. We call $\xi_m$ the "modal purity" because it is the probability that the $m$ or more observed photons were actually subtracted from the mode matched to the local oscillator.

In practice we calculate these states and their associated probabilities by use of their Wigner functions, according to the methods described in [7]. Eventually we obtain $W(q, p)$, the Wigner function for $\tau_1(m)$, which depends on $V_0$, $\eta_s$, $R$, $\mu$, $\eta_h$, the $\xi_n$'s, and $m$. We then perform a least squares fit of these Wigner functions to the Wigner functions reconstructed by maximum likelihood from the homodyne measurements to obtain estimates for the parameters $V_0$, $\eta_s$ and the $\xi_n$'s. From separate measurements we know $R$, $\mu$ and $\eta_h$, and fix these parameters in the fitting routine. A package to calculate the Wigner functions based on the above model can be obtained from the authors upon request.

## III. Experimental findings

We have used the above model to fit our experimental data and obtain the experimental parameters. Figure A3 shows the Wigner functions obtained by maximum likelihood reconstruction from the homodyne data. We used a maximum number of 20 photons for the reconstruction algorithm. Figure A3 also shows Wigner functions obtained by fitting our model to the maximum likelihood reconstructions. Table I lists our experimental findings. The beam splitter reflectivities were adjusted from 2.5 % to 20 % in the different photon-number-subtraction experiments. Higher fidelities are predicted for lower reflectivities, but we increased the reflectivity to increase the frequency of higher photon number subtraction events. For the fitting routine, the overall detection efficiency $\mu$ was set to 0.17 for the TES and to 0.08 for the APD measurements. Inferring other experimental parameters by fitting the model described above to the one photon subtraction measurements, we find $\eta_s = 0.72$, $V_0 = 0.229$ (-6.4 dB), $V_q = 3.423$ (+5.3 dB), $V_p = 0.445$ (-3.5 dB) and $\xi_1 = 0.91$.

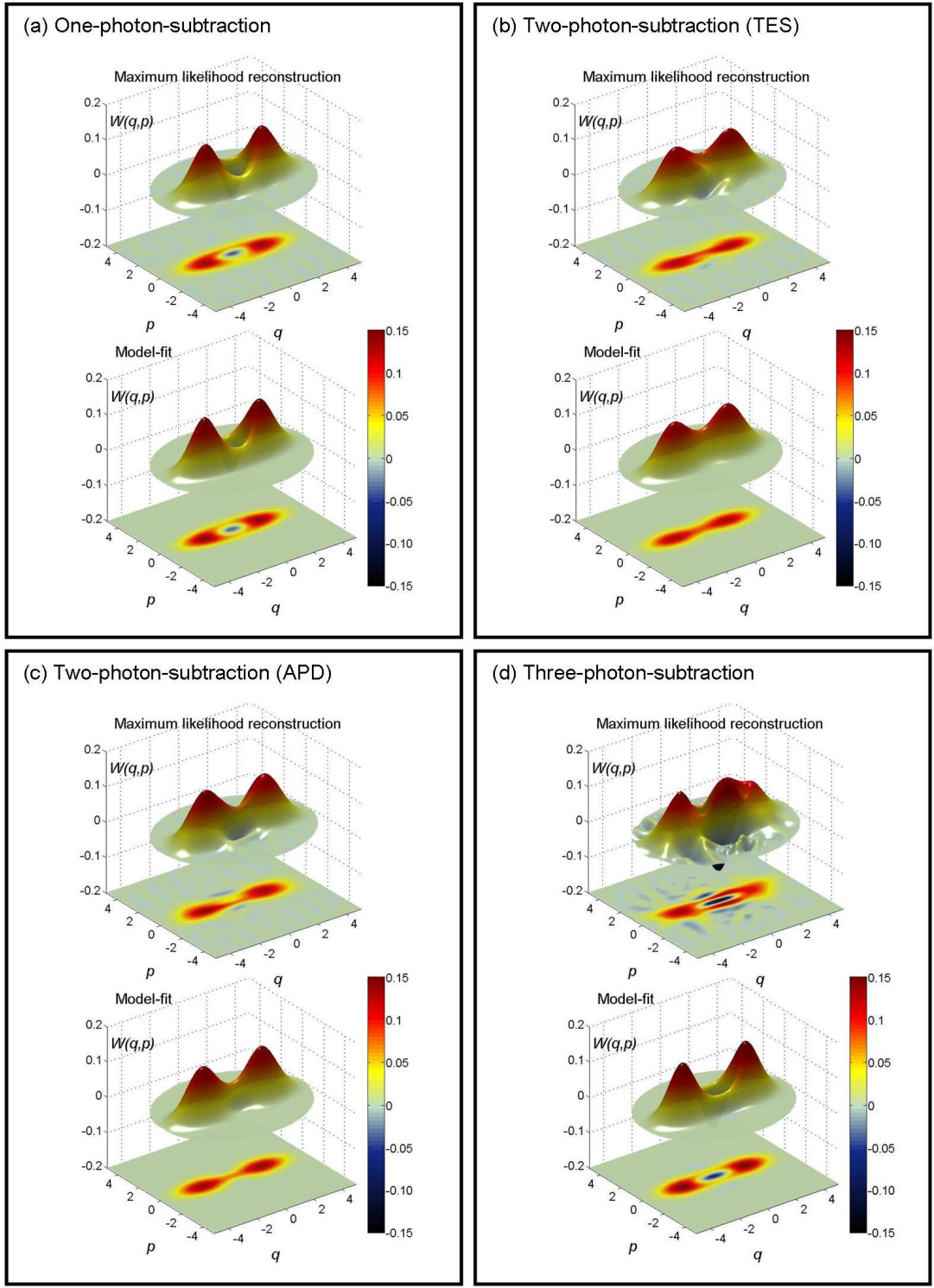

FIG. A3. | Wigner function plots of CSSs created in our experiments. (a)-(d) One, Two and Three photon subtraction maximum likelihood reconstruction and model-fit, respectively.

|  | one photon subtraction (APD) | two photon subtraction (TES) | two photon subtraction (APD) | three photon subtraction (TES) |
|---|---|---|---|---|
| R | 2.5% | 10% | 10% | 20% |
| $\eta_s$ | 0.72 | 0.71 | 0.72 | 1 |
| $V_0$ | 0.23 | 0.24 | 0.24 | 0.36 |
| $\xi_0$ | 0.09 | 0.20 | 0.06 | 0.00 |
| $\xi_1$ | 0.91 | 0.18 | 0.09 | 0.01 |
| $\xi_2$ | --- | 0.62 | 0.85 | 0.15 |
| $\xi_3$ | --- | --- | --- | 0.84 |
| $\lvert\alpha\rvert$ | $1.32^{+0.01}_{-0.02}$ | $1.16^{+0.04}_{-0.04}$ | $1.30^{+0.04}_{-0.02}$ | $1.76^{+0.02}_{-0.19}$ |
| $\langle n\rangle_\alpha$ | 1.75 | 1.17 | 1.58 | 3.08 |
| $\mathscr{F}$ | $0.522^{+0.004}_{-0.010}$ | $0.531^{+0.017}_{-0.018}$ | $0.523^{+0.022}_{-0.014}$ | $0.592^{+0.036}_{-0.142}$ |
| $\langle n\rangle_\rho$ | $1.96^{+0.05}_{-0.04}$ | $1.89^{+0.05}_{-0.06}$ | $2.34^{+0.06}_{-0.05}$ | $2.75^{+0.06}_{-0.24}$ |
| $\mathscr{W}^\rho_{\min}$ | $-0.041^{+0.009}_{-0.001}$ | $-0.010^{+0.001}_{-0.001}$ | $-0.018^{+0.002}_{-0.002}$ | $-0.116^{+0.073}_{-0.019}$ |
| datapoints | 324,000 | 25,000 | 39,000 | 1087 |
| integration time | ~3 hours | ~24 hours | ~120 hours | ~60 hours |

Table I Experimental findings. $V_0$, $\eta_s$ and $\xi_n$ were obtained from a least squares fit of the above model. $\alpha$, $\langle n\rangle_\alpha$ and $\mathscr{F}$ were obtained by comparing the maximum likelihood state estimate with a theoretical CSS that gave highest fidelity. $\langle n\rangle_\rho$ is the average photon number in the reconstructed state. $\mathscr{W}^\rho_{\min}$ is the minimum of the reconstructed Wigner function. The photon subtraction beam splitter reflectivity R was determined by a separate measurement.

The model finds a better fit using a higher squeezing purity than is directly measured with homodyne detection. This may be attributed to using our single-mode model to describe our multi-mode states. A thorough investigation based on the multi-mode model in [6] may clarify this discrepancy. The fits to the data reveal a constant squeezing purity of about $\eta_s = 0.72$ for all one and two photon subtraction experiments. When we fit the model to the three photon subtraction data, we find that a squeezing purity of $\eta_s = 1$ provides the best fit. However, we know from other measurements that $\eta_s < 0.75$. The failure of the model in the three photon subtraction experiment may be caused by numerical difficulty obtaining the correct fit and/or multimode effects [6]. The modal purity $\xi_m$ of the subtracted photons is at least 0.84, except for the two photon TES experiment, where spurious LO photons scattered into the TES contribute to 38% false heralds, as determined from the modal purity. The scattered LO photons arrive 5 ns after the true signal photons; the delay is determined by the beam paths. Therefore, gating the APD with a window smaller than 5 ns suppresses the spurious LO contribution. This gating is possible only because of the APD's small jitter (≈400 ps). The TES (jitter ≈ 100 ns) does not allow for such accurate gating. Therefore the modal purity of the subtracted photons is lower in the TES case. Note that the lower modal purity could be improved. In our case the spurious photons originate from a reflection off one output port of the polarizing beam splitter that combines the CSS and the LO (PBS$_3$ in figure A1). For example, we could use a slightly wedged output port surface which would ensure the reflection into a spatial mode that is orthogonal to the subtraction arm's spatial mode.

The reason for the higher modal purity in the three photon subtraction experiment is that after increasing the subtracting beam splitter's reflectivity, the rate of subtracting three "good" photons was increased while the rate of detecting scattered LO photons was decreased. When reporting the fidelity of the states produced in our experiment, we maximize the fidelity over all ideal CSSs, obtaining the amplitude $\alpha$ of the highest fidelity CSS. The mean photon number $\langle n \rangle_\alpha$ of that CSS is calculated via

$$\langle n \rangle_\alpha = |\alpha|^2 \frac{\exp\left(2|\alpha|^2\right) - \cos(\varphi)}{\exp\left(2|\alpha|^2\right) + \cos(\varphi)}, \tag{A13}$$

where $\phi = 0$ for an even CSS and $\phi = \pi$ for an odd CSS.



\end{document}